# Extensive air showers without muon component

A.A. Mikhailov, N.N. Efremov, E.S. Nikiforova.
*Yu.G. Shafer Institute of Cosmophysical Research and Aeronomy, 31 Lenin Ave., 677980 Yakutsk, Russia*
Presenter: A. Mikhailov (mikhailov@ikfia.ysn.ru), rus-mikhailov-AA-abs3-HE14-poster

Extensive air showers of cosmic rays with energies above $5 \times 10^{18}$ eV registered by the Yakutsk array have been analyzed. The existence of showers without muon component is found. Among them we found 5 clusters and these clusters correlate with nearest pulsars. It is found that there exist 4 classes of EAS'. Showers with high content of muons are mainly observed at $E > 4.10^{19}$ eV and they are formed most likely by heavy nuclei.

## 1. Introduction

We consider a portion of muons in extensive air showers (EAS') by using Yakutsk array data. The showers with $E > 5.10^{18}$ eV, zenith angles $< 60°$ and cores inside of the array perimeter are analyzed. Previously, by Yakutsk and AGASA array data the showers without muons were detected [1-3].

## 2. Discussion

All registration time at the Yakutsk array was divided into separate periods with 6 h intervals. The time period, when the muon detector did not operate, was excluded from the analysis. At first, the showers without muons were selected (energy threshold of muon EAS registration with detectors is 1 GeV). When the muon detector observation is zero, the probability that the given detector does not operate is estimated by a formula $P = \Pi(P_{1i} + P_{2i})$, where $P_{1i}$ is the probability that neither of shower particles does not fall into a detector, $P_{2i}$ is the probability that one particle falls into a detector and it does not operate. If the probability $P > 10^{-3}$ then the given EAS is excluded from consideration.

Fig.1a presents the particle density by data of scintillation detectors which register electron-photon EAS component (circles) depending on the distance to a shower core. The EAS with $E \sim 2 \times 10^{19}$ eV and zenith angle $\theta = 18.1°$, registered on December 8, 1994, is shown. The muon EAS component is equal to zero (open circles) at the five muon stations. The probability that neither of particles does not fall into the 5 muon detectors of 100 m$^2$ total area is $\sim 10^{-20}$. The electron-photon component densities expected are presented by a solid line. Every case of registration of a shower without muons is carefully controlled. A EAS without muons is selected when during 30 min before and after of the given event the muon detectors operate from the particles of another EAS [2].

Fig.2 shows the distribution of 19 EAS' with $E > 5.10^{18}$ eV without muons (black circles) in the equatorial coordinate system ($\delta$ - declination, RA – right ascension). In the EAS distribution, one cluster with 4 showers and 4 doublets are observed (see Table and Fig.1). The last doublet consists of EAS' without muons and deficient by muons. The distance between particles in the doublets is $<9°$. The arrival directions of all 4 EAS' of the first cluster (Table, Fig.1) are inside of 9° from the pulsar PSR 0809+74 [4], which is spaced at the distance of 0.3 kpc from the Earth. The probability of chance that shower directions of 4 EAS' from 19 are inside 9° from the pulsar PSR 0809+74 is $\sim 5.10^{-3}$. The arrival directions of 2-4 EAS' doublets are inside 9° from the pulsars PSR 0450+55, 2217+47, 2241+65 [4], respectively, which are located at the distance less than 2.4 kpc from the Earth (Table 1). Inside the distance of 9° from the doublet 5 a pulsar is not found. At $E>4 \cdot 10^{19}$ eV neither shower without muons is not detected. A portion of EAS' without muons to the total number of EAS' is ~ 1%.

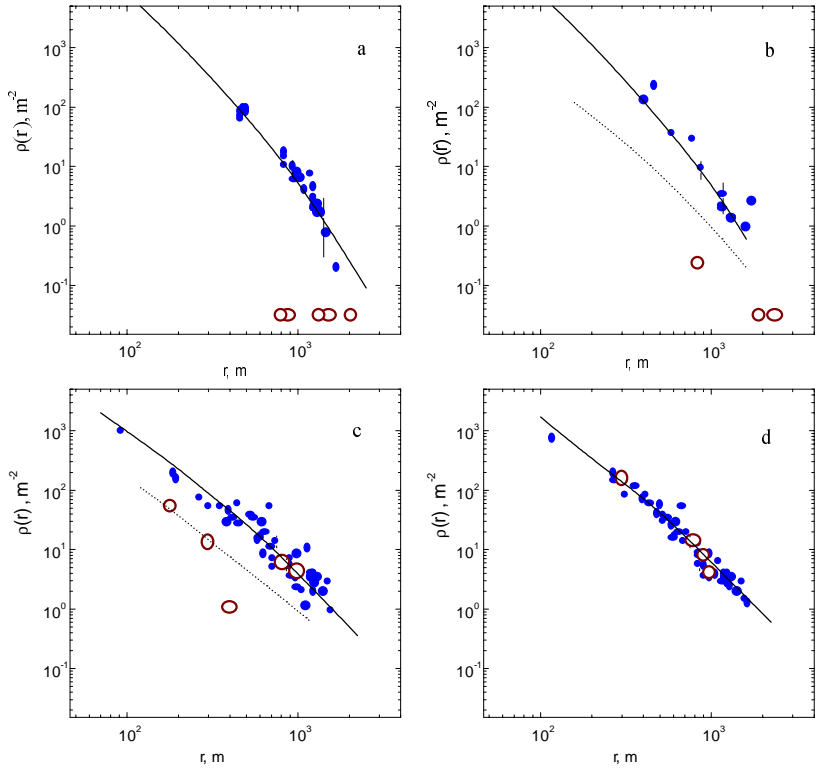

Fig.1. a) EAS' without muons, b) EAS' with poor by muons, c) usual EAS', d) EAS' with large number of muons. ● - scintillation detectors, ○ - muon detectors. Fig.1a shows that muon EAS component for 5 detectors is zero. The expected densities of electron-photon component (solid) and muon (dashed line) versus the distance r from a shower core.

**Table 1.** Arrival time, energy and coordinates of the showers forming the clusters and pulsars which correlate with them

| No of cluster | Date | Energy, EeV | Right ascension and declination of showers, degs. | | Pulsars, PSR | Distance, kpc | LgT, years |
|---|---|---|---|---|---|---|---|
| 1 | 27.01.1992 | 13.3 | 148.3 | 70.0 | 0809+74 | 0.3 | 8.1 |
| 1 | 18.03.1994 | 47.4 | 92.5 | 74.0 | | | |
| 1 | 08.12.1994 | 21.3 | 123.1 | 76.5 | | | |
| 1 | 10.05.2001 | 7.1 | 136.2 | 70.0 | | | |
| 2 | 25.04.1996 | 8.6 | 81.7 | 60.2 | 0450+55 | 0.7 | 6.3 |
| 2 | 26.03.1998 | 8.3 | 85.4 | 57.6 | | | |
| 3 | 17.11.2000 | 5.7 | 322.4 | 51.4 | 2217+47 | 2.4 | 6.5 |
| 3 | 13.01.2002 | 7.7 | 326.7 | 44.3 | | | |
| 4 | 13.03.1992 | 8.7 | 334.7 | 69.8 | 2241+69 | 2.3 | 6.7 |
| 4 | 30.05.1996 | 5.0 | 322.9 | 65.8 | | | |
| 5 | 25.04.1996 | 8.6 | 175.7 | 64.1 | No pulsars <9° | | |
| 5 | 29.12.1998 | 6.4 | 172.2 | 66.4 | | | |

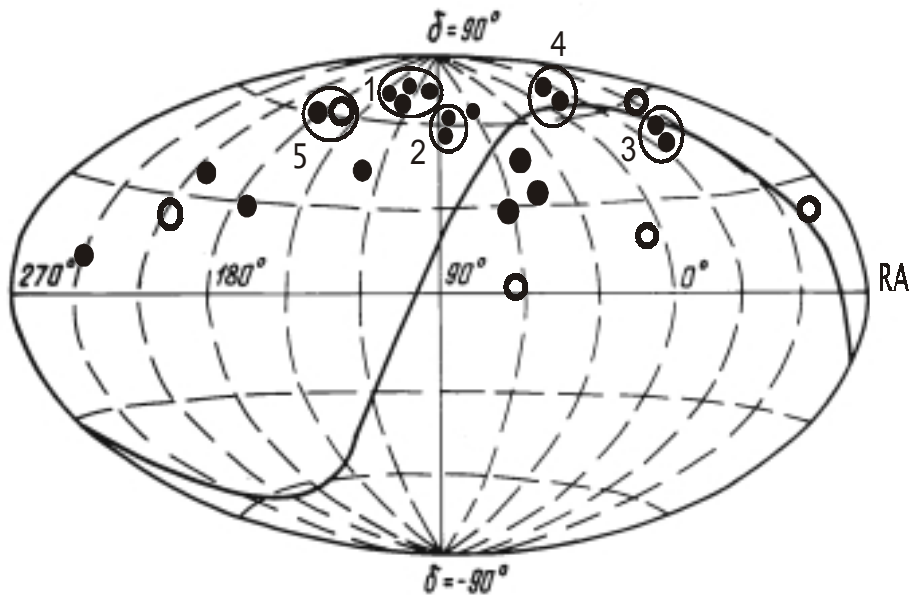

Fig.2. Distribution of showers without muons (black circles) and poor by muons (open circles) in the equatorial coordinate system δ and RA. δ - declination, RA – right ascension. The large circles are clusters 1 - 5.

We found 6 EAS' with $E>5 \cdot 10^{18}$ eV whose muon density deficit more 3σ of than it is expected from the usual showers. For example, the shower on April 24, 1991 (Fig.1b) has the muon density 4.7 part./detector at the distance 851 m from the shower core, that is small by 3.2σ than it is expected according [5] from the typical shower (28.1 part./detector). The distribution of 6 EAS' with $E>8 \cdot 10^{18}$ eV deficient by muons on the celestial sphere is shown in Fig.2 (open circles). Their portion of the total number of showers is ~ 1%.

Note that the distribution of EAS' without and pour muons are isotropic. They are correlate with pulsars independently position of pulsars (Fig.2). At same time EAS' with typical content of muons correlate with pulsars only situated along force lines of magnetic field of Local Arm [6,7]. Probably these EAS' are formed by a charged particles, EAS' without and pour muons are formed by neutral particles.

4 classes of showers are found:
1) showers without muon component inside the limits of the registration threshold of detectors (Fig.1a) - ~1%,
2) showers deficient by muons (Fig.1b) – ~1%,
3) showers with typical content of muons (Fig.1c) - ~97%,
4) showers with high content of muons (Fig.1d) - <0.01%.

Fig.1d presents the shower on May 7, 1989 with $E=1.2 \times 10^{19}$ eV, the maximum energy observed at the Yakutsk array consisting of the muon only. The showers with high content of muons are mainly observed at the highest energies, $E>4.10^{19}$ eV [5]. Therefore EAS' with high content of muons are formed most likely by heavy nuclei [8].

## 3. Conclusions

The ultrahigh energy EAS' without muon component are detected. The arrival directions of certain of them are closely related to each other and 5 clusters are found. The arrival directions for 4 clusters of 5 correlate with nearest pulsars. It is shown that there exist 4 classes of EAS. The showers with the large number of muons are only observed at the highest energies. Possible cosmic rays at $E> 4.10^{19}$ eV are heavy nuclei.

## 4. Acknowledgements

The work has been supported by RFBR (project N 04-02-16287).The Yakutsk EAS array was supported by the Ministry of Training of the Russian Federation, project no.01-30.

## References

[1]  A.A. Mikhailov, E.S. Nikiforova, JETF Lett. 72, 229 (2000).
[2]  A.A. Mikhailov, E.S. Nikiforova, Proc. 27th ICRC, Hamburg (2001) 1, 417.
[3]  K. Shinozaki, M. Teshima, GZK and surroundings. Ed. by C. Aramo, Catania 18 (2004).
[4]  N. Taylor et al., Astrophys. J. Suppl. 88**,** 529 (1993).
[5]  A.V. Glushkov et al., Astropart. Phys. 4, 15 (1995).
[6]  A.A. Mikhailov. Nucl.Phys.B. Proc.Suppl. 75A, 359 (1999).
[7]  A.A. Mikhailov, JETF Lett. **77**, 151 (2003).
[8]  G.B. Khristiansen et al., Kosm. Izluch. Sverch. En. M., 1975.